\documentstyle[pmart,twoside,fleqn,epsf]{article} % Do not change this line!

\begin{document}

\title{  Spin--Orbital Physics in Transition Metal Oxides }

\author{ Andrzej M. Ole\'s }

\address{Marian Smoluchowski Institute of Physics, Jagellonian
         University, \\ Reymonta 4, PL--30059 Krak\'ow, Poland\\
         Max--Planck--Institut f\"ur Festk\"orperforschung,\\
         Heisenbergstrasse 1, D--70569 Stuttgart, Germany}

\date{September 22, 2008}

\maketitle


\begin{abstract}
We present the main features of the spin-orbital superexchange
which describes the magnetic and optical properties of Mott
insulators with orbital degrees of freedom. In contrast to the
SU(2) symmetry of spin superexchange, the orbital part of the
superexchange obeys the lower cubic symmetry of the lattice and is
intrinsically frustrated. This intrinsic frustration and
spin-orbital entanglement induce enhanced quantum fluctuations,
and we point out a few situations where this leads to disordered
states. Strong coupling between the spin and orbital degrees of
freedom is discussed on the example of the $R$VO$_3$ perovskites,
with $R$ standing for rare-earth ion, La,$\cdots$,Lu. We explain
the observed evolution of the orbital $T_{\rm OO}$ and N\'eel
$T_{N1}$ transition temperature in the $R$VO$_3$ series with
decreasing ionic radius $r_R$. A few open problems and the current
directions of research in the field of spin-orbital physics are
pointed out. \\
{\it Published in Acta Phys. Polon.} A {\bf 115}, 36 (2009)
\end{abstract}

\section{Orbital versus spin superexchange}
\label{sec:orb}

In recent years
the physical properties of Mott (or charge transfer) insulators are
in the focus of interest in the condensed matter theory \cite{Ima98}.
In order to develop theoretical understanding of complex phenomena in doped
correlated insulators, including high temperature superconductivity in the
cuprates and the colossal magnetoresistance (CMR) in the manganites, it is
necessary to describe first the undoped materials, such as La$_2$CuO$_4$
and LaMnO$_3$. In both cases the local Coulomb interactions (Hubbard $U$)
is large and suppresses charge fluctuations, leading to low-energy
effective Hamiltonians with superexchange interactions which stabilize
antiferromagnetic (AF) spin order at low temperature \cite{Kug82,Cnr78}.
However, these two compounds are qualitatively quite different. On the one
hand, the degeneracy of
partly filled $e_g$ orbitals is lifted in La$_2$CuO$_4$ by the tetragonal
distortions of CuO$_6$ octahedra, resulting in two-dimensional (2D) AF
superexchange of $S=1/2$ holes in $x^2-y^2$ orbitals of Cu$^{2+}$ ions.
On the other hand, in LaMnO$_3$ $e_g$ orbital degeneracy plays
a fundamental role and, together with the Jahn-Teller (JT) distortions,
is required to understand the origin of the anisotropic $A$-type AF
($A$-AF) order \cite{Fei99}. As realized already three
decades ago \cite{Kug82},
in this latter case the orbital degrees of freedom,
which are described by the components
\begin{equation}
|x\rangle \equiv \frac{1}{\sqrt{2}}\left(x^2-y^2\right),\hskip 1cm
|z\rangle \equiv \frac{1}{\sqrt{2}}\left(3z^2-r^2\right)
\label{generic}
\end{equation}
of pseudospin $\tau=1/2$ and contribute explicitly
to the structure of the superexchange. Thus they have to be included
on equal footing with ionic spins in a spin-orbital superexchange model.
In the last two decades several new
concepts were developed, such as enhanced quantum fluctuations due to
orbital degrees of freedom which participate in joint spin-orbital
excitations \cite{Fei97}, and spin-orbital entanglement which occurs in
cases when spin and orbital operators cannot be decoupled from each other
\cite{Ole06}. The actual physical problems in this emerging and
rapidly developing field were reviewed in the Focus
Issue {\it Orbital Physics\/} in New Journal of Physics \cite{NJP}
a few years ago. Here we want to focus on a few representative
recent developments in the field of spin-orbital physics.

Before we will discuss a few consequences of realistic spin-orbital models,
let us consider first purely orbital superexchange interactions, as
realized in ferromagnetic (FM) 2D systems, such as K$_2$CuF$_4$, or in
FM $ab$ planes of three-dimensional (3D) perovskites, either in KCuF$_3$
or in LaMnO$_3$. The superexchange originates from
charge $d_i^9d_j^9\rightleftharpoons d_i^8d_j^{10}$ excitations
(with $d_i^9\equiv t_{2g}^6e_g^3$, $d_i^8\equiv t_{2g}^6e_g^2$) in
the first case (for Cu$^{2+}$ ions), and from
$d_i^4d_j^4\rightleftharpoons d_i^3d_j^5$ (with
$d_i^4\equiv t_{2g}^3e_g$, $d_i^3\equiv t_{2g}^3$, and
$d_i^5\equiv t_{2g}^3e_g^2$) excitations in the second one
(for Mn$^{3+}$ ions) \cite{Ole05}.
These charge excitations can be transformed away by applying a similar
canonical transformation to the one leading to the $t$-$J$ model
\cite{Cha77}, commonly used to describe
the superconducting cuprates. For the FM compound only high-spin
excited states contribute, and in the strong-coupling regime ($t\ll U$)
one obtains the effective low-energy Hamiltonian describing orbital
superexchange between pseudospins $\tau=1/2$ due to $e_g$ electrons
(holes in KCuF$_3$) in a cubic crystal \cite{Fei97,vdB99},
\begin{equation}
{\cal H}_o = H_J + H_z\,.
\label{orbis}
\end{equation}
the first term $H_J$ in Eq. (\ref{orbis}) describes superexchange
interaction which depends on the superexchange constant
\begin{equation}
J=\frac{4t^2}{U}
\label{sex}
\end{equation}
[here $t$ is
the $(dd\sigma)$ hopping element between two $|z\rangle$ orbitals
along the $c$ axis, and $U$ is the intraorbital Coulomb interaction],
and on Hund's exchange element
\begin{equation}
\eta = \frac{J_H}{U}\,.
\label{eta}
\end{equation}
This interaction parametrizes the multiplet structure of transition
metal ions when anisotropy of Coulomb and exchange elements is neglected
\cite{Ole05}. In the simplest case of $d^9$ system, the singlet-triplet
splitting between the two lowest $d^8$ excited states is $2\eta$
\cite{Ole00}.

Following Ref. \cite{vdB99}, we will consider below a 3D (perovskite)
system with FM spin order and orbital interactions described by the
Hamiltonian (\ref{orbis}). The origin of intrinsic frustration in the
orbital superexchange is best realized by considering its form ---
\begin{equation}
H_J = {1\over 2}J r_1 \sum_{\langle ij\rangle} \left(
  \tau^{(\gamma)}_i\tau^{(\gamma)}_j-{1\over 4}\right)\,,
\label{HJ0}
\end{equation}
where $\gamma=a,b,c$ refers to the cubic axes, and $r_1=1/(1-3\eta)$
follows from the energy of the high-spin charge excitation \cite{Ole05}.
The pseudospin operators are defined as follows,
\begin{equation}
\tau^{(a,b)}_i = {1\over 4}\left( -\sigma^z_i \pm \sqrt{3}\sigma^x_i
                           \right), \hskip 1cm
\tau^{(c)}_i   = {1\over 2}   \sigma^z_i,
\label{eg}
\end{equation}
where $\sigma^{x(z)}_i$ are Pauli matrices and sign $+$ ($-$) is selected
for a bond $\langle ij\rangle$ along $a$ ($b$) axis.
The pseudospin interactions
$\propto\tau^{(\alpha)}_i\tau^{(\alpha)}_j$ favor alternating orbitals (AOs)
on each bond. They are fundamentally different from the SU(2)-symmetric
spin interactions, as they: ($i$) obey only lower cubic symmetry,
($ii$) are Ising-like, having only one component of the pseudospin
interaction which favors pairs of orbitals oriented along the bond
($z$-like) and in the plane perpendicular to the bond ($x$-like), and
($iii$) change their form when the cubic direction is changed. The
interactions (\ref{HJ0}) look to be classical, but in fact due to the
form of of the pseudospin $e_g$ operators (\ref{eg}) they are not.
However, the quantum corrections generated by them are rather small
\cite{vdB99}.

The second term $H_z$ in Eq. (\ref{orbis}) stands for the orbital splitting
(by $J\varepsilon_z$), and follows from finite crystal field at either
Cu$^{2+}$ or Mn$^{3+}$ ions,
\begin{equation}
H_z =  - J \varepsilon_z \sum_i \tau^{(c)}_i.
\label{Hz}
\end{equation}
The crystal-field splitting favors electron (hole) occupancy in $x$
($z$) orbital for $\varepsilon_z>0$ or ($\varepsilon_z<0$). This term
is induced by static distortions in the tetragonal field and is of
particular importance in 2D systems, e.g. in the high-$T_c$ cuprates.

The directional nature of pseudospin orbital interactions is responsible
for their intrinsic frustration \cite{Fei97,vdB04}. In fact, the pair of
orbitals which would minimize the energy for a bond $\langle ij\rangle$
is different for each cubic axis. Therefore, while for a single bond the
minimal energy of $-\frac{1}{4}J$ can easily be obtained by selecting a
pair of orthogonal orbitals on both sites, such as $|z\rangle$ and
$|x\rangle$ orbital for a bond along the $c$ axis (note that this
corresponds to the Ising superexchange interaction as only one orbital
is active in intersite charge excitations), this is no longer possible
for a 2D (or 3D) system. Thus, unlike in spin systems, the tendency
towards disorder (quantum {\it orbital liquid\/}) is enhanced with
increasing system dimension \cite{Fei05}.

The essence of orbital frustration which characterizes the $e_g$ orbital
superexchange (\ref{HJ0}) is captured by the 2D compass model,
originally developed as a model for Mott insulators \cite{Kug82}.
Intersite interactions in the compass model are descibed by products
$\tau^{\alpha}_i\tau^{\alpha}_j$ of pseudospin components,
where $\alpha=x,y,z$, rather than by a scalar product
${\vec \tau}_i\cdot{\vec \tau}_j$.
\begin{equation}
\tau^x_i = \frac{1}{2}\sigma^{x}_i , \hskip 1cm
\tau^y_i = \frac{1}{2}\sigma^{y}_i , \hskip 1cm
\tau^z_i = \frac{1}{2}\sigma^{z}_i ,
\label{t2g}
\end{equation}
where $\alpha=x,y,z$, rather than by a scalar product
${\vec \tau}_i\cdot{\vec \tau}_j$. In the 2D case the
$\tau^x_i\tau^x_j$ interactions for bonds $\langle ij\rangle$ along the
$a$ axis compete with the $\tau^z_i\tau^z_j$ ones along the $b$ axis
\cite{Kho03}.
Recently certain aspects of this model were investigated by analytic
\cite{Kho03} and numerical \cite{Mil05,Wen08} methods. Despite its
closeness to ordinary models used in quantum magnetism, there is no
ordered phase with finite magnetization. Thus, this case is qualitatively
different from the frustrated Ising interactions on the 2D square lattice,
where it was proven rigorously that magnetic order develops below certain
transition temperature which depends on the ratio of frustrated to
nonfrustrated plaquettes \cite{Lon80}. A competition of pseudospin
interactions along different directions results instead in intersite
correlations similar to the anisotropic XY model,
and competition between two types of Ising-like
order generates a quantum phase transition with high degeneracy of the
ground state when all interactions have the same strength \cite{Mil05}.

It is interesting to note that a similar quantum phase
transition exists also in the one-dimensional (1D) compass model
\cite{Brz07}, when both above interactions alternate along the chain
($N'=N/2$ is the number of unit cells):
\begin{equation}
{\cal H}_{1D}=\sum_{i=1}^{N'}
\left\{ J_x\tau_{2i-1}^x \tau_{2i}^x +
        J_z\tau_{2i}^z   \tau_{2i+1}^z \right\}\,.
\label{H1}
\end{equation}
This model was solved exactly in the whole range of $\{J_x,J_z\}$
parameters \cite{Brz07} by maping to the exactly solvable quantum
Ising model \cite{Perk} in different subspaces (see also this volume
\cite{Brz08}). Equal coupling constants $J_x=J_z=J$ correspond to the
the quantum critical point, where the orbital liquid emerges
from two different disordered phases, with hidden order of pairs
of pseudospins on every second bond. Presented here
1D model model (\ref{H1}) provides thus a beautiful example
of a first order quantum phase transition.

\section{Spin-orbital models and exotic liquid states}
\label{sec:som}

The structure of spin-orbital superexchange
involves interactions between SU(2)-symmetric spin scalar products
${\vec S}_i\cdot{\vec S}_j$ on each bond $\langle ij\rangle$ connecting
two nearest-neighbor transition metal ions,
each one coupled to orbital operators $\{{\vec \tau}_i,{\vec \tau}_j\}$
which obey only much lower symmetry (at most cubic for a cubic lattice)
and its general form is \cite{Ole05},
\begin{equation}
{\cal H}_J = J \sum_{\langle ij \rangle } \left\{
{\hat {\cal J}}_{ij}^{(\gamma)} \left( {\vec S}_i \cdot {\vec S}_j +S^2\right) +
{\hat {\cal K}}_{ij}^{(\gamma)} \right\}.
\label{som}
\end{equation}
The orbital operators ${\hat{\cal J}}_{ij}^{(\gamma)}$ and
${\hat{\cal K}}_{ij}^{(\gamma)}$ involve the active orbitals on each
bond $\langle ij \rangle$ along  direction $\gamma$, either $e_{g}$ or
$t_{2g}$, which participate in $d^n_i d^n_j \rightleftharpoons
d^{n+1}_i d^{n-1}_j$ virtual excitations, and thus have the symmetry of
the lattice ({\it e.g.}~cubic in the perovskites). The superexchange model
(\ref{som}) consists of several terms which originate from different
charge excitations. This feature made it possible to relate the averages
of these different excitations to the spectral weights in the optical
spectroscopy
\cite{Kha04}, and this serves now as a theoretical tool to explain
the observed anisotropy and temperature dependence of the spectral weights
in the optical spectra \cite{Ole05}.

The best known spin-orbital model is the Kugel-Khomskii model for $d^9$
Cu$^{2+}$ ions which involves the $e_g$ orbital operators (\ref{eg}) and
includes both FM and AF spin interactions, depending on whether the
above charge excitation (for $n=9$) involves the high-spin (triplet) or
one of the low-spin (singlet) $d^8$ states. These excitations
are parametrized by the coefficients $r_1=1/(1-3\eta)$, $r_2=1/(1-\eta)$
and $r_4=1/(1+\eta)$. Using compact
notation of Ref. \cite{Ole00} it can be written as follows:
\begin{eqnarray}
{\cal H}_J(d^9)\!&=&\!\frac{1}{2}J\sum_{\langle ij\rangle}
\left\{\left[-r_1\left(\vec{S}_i\!\cdot\!\vec{S}_j+\frac{3}{4}\right)
+r_2\left(\vec{S}_i\!\cdot\!\vec{S}_j-\frac{1}{4}\right)\right]
\left(\frac{1}{4}-\!\tau_i^{(\gamma)}\tau_j^{(\gamma)}\right)\right.\nonumber \\
& &\left. \hskip .5cm +(r_2+r_4)
\left(\frac{1}{4}-\!\tau_i^{(\gamma)}\tau_j^{(\gamma)}\right)
\left(\frac{1}{2}-\!\tau_i^{(\gamma)}\right)
\left(\frac{1}{2}-\!\tau_j^{(\gamma)}\right)\right\}.
\label{som9}
\end{eqnarray}

In ordeer to derive magnetic excitations for the systems with orbital
degeneracy one usually derives magnetic exchange constants for a bond
$\langle ij \rangle$ by averaging over the orbital operators in Eq.
(\ref{som}),
\begin{equation}
J_{ij}=\langle {\hat{\cal J}}_{ij}^{(\gamma)}\rangle.
\label{Jij}
\end{equation}
This procedure assumes implicitly that spin and orbital
operators can be decoupled from each other and ignores the possibility
of entanglement and composite spin-orbital excitations introduced in Ref.
\cite{Fei97}. It turns out that such excitations play a prominent role in
destabilizing the classical AF long-range order in the $d^9$ spin-orbital
model \cite{Fei98}.

Qualitatively, the classical phase diagram of model (\ref{som9})
including the crystal-field term (\ref{Hz}) consists in the
$\{\varepsilon_z,\eta\}$ plane of four AF phases shown schematically in
Fig. 1. First, if $\eta$ is small, then:
($i$) if $\varepsilon_z>0$, the order is AF and the holes occupy
$|x\rangle$ orbitals (2D AFx phase), while
($ii$) for $\varepsilon_z<0$ --- $|z\rangle$ orbitals are occupied in
the anisotropic 3D AFz phase.
Second, if $J_H/U$ is large the high-spin excitations become important
for the bonds with different orbitals on both sites, and weak FM
interactions ($\propto\eta J$) occur within the $ab$ planes, accompanied
by strong ($\propto J$) AF interactions along the $c$-axis, leading to
$A$-AF phase. Depending on the actual value of $\varepsilon_z$, either
($iii$) $A$-AF1 phase, or
($iv$) $A$-AF2 phase [with AF $bc$ planes and FM bonds along the $a$ axis]
is stable on the classical level. The energies of all the classical
phases are degenerate at the $(\varepsilon_z,\eta)=(0,0)$ point. The
competition between the above types of spin and orbital order leads
here to the quantum critical point at $(\varepsilon_z,\eta)=(0,0)$.

%%%%%%%%%%%%%%%%%%%%%%%%%%%%%%%%%%%%%%%%%%%%%%%%%%%%%%%%%%%%
%%                      figure 1
%%%%%%%%%%%%%%%%%%%%%%%%%%%%%%%%%%%%%%%%%%%%%%%%%%%%%%%%%%%%
\begin{figure}[t!]
  \epsfysize=6cm
  \centerline{\epsfbox{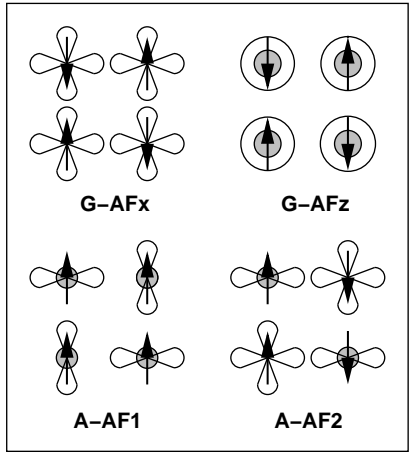}}
\caption
{Schematic representation of orbital and magnetic long-range order in
the $ab$ plane of the $d^9$ model (\ref{som9}) in: $G$-AFx, $G$-AFz,
$A$-AF1, and $A$-AF2 phase. The shadded parts of
different orbitals are oriented along the $c$-axis. The spins (arrows) in
the next $ab$ plane along the $c$ axis are AF to those below them in both
$G$-AFz and $A$-AF1 phase, and FM in the
$A$-AF2 phase. In the $G$-AFx phase there is no
superexchange coupling along the $c$ axis to the next plane, but
this degeneracy is easily removed when
$|x\rangle$ orbitals mix weakly with $|z\rangle$ orbitals. }
\label{fig:mfa}
\end{figure}

It has been argued \cite{Fei97,Fei98} that, similar to the situation
in frustrated $J_1$-$J_2$-$J_3$ spin models in quantum magnetism where
the quantum corrections obtained from the spin-wave theory diverge and
suppress the long-range order near quantum critical points \cite{Cha88},
the spin order in all the $G$-AF and $A$-AF phases is suppressed near
the $(\varepsilon_z/J,\eta)=(0,0)$ point. Although couterarguments were
also given that the AF$z$ (but not AF$x$!) phase could be stabilized by
order-out-of-disorder mechanism \cite{Kha98} and this controversial
issue awaits still an answer with a more sophisticated treatment, it is
believed that only short-range order with correlations of valence-bond
(VB) type may survive near the transitions between different phases with
classical order (Fig. 1). In Ref. \cite{Fei97} a few representative
variational VB wave functions (see Fig. 2) were constructed, with spin
singlets on individual bonds $\{\langle ij\rangle\}$ accompanied by the
variationally optimized orbitals pointing predominantly along the bond.
Each VB phase is characterized by a set $K$ which defines the bonds
$\{\langle ij\rangle\}$ occupied by spin singlets, covering the 3D lattice.
The two simplest (dimer) VB states are (see Fig. 2):
 ($i$) singlets along the $a$ axis with orbitals close to $3x^2-r^2$
       (VB$a$ phase, degenerate with the analogous VB$b$ phase),
($ii$) singlets along the $c$-axis with $|z\rangle$ orbitals (VB$c$),
are indeed more stable than the above classical AF phases, and we
emphasize that this happens in three dimensions.

%%%%%%%%%%%%%%%%%%%%%%%%%%%%%%%%%%%%%%%%%%%%%%%%%%%%%%%%%%%%
%%                      figure 2
%%%%%%%%%%%%%%%%%%%%%%%%%%%%%%%%%%%%%%%%%%%%%%%%%%%%%%%%%%%%
\begin{figure}[t!]
  \epsfysize=7cm
  \centerline{\epsfbox{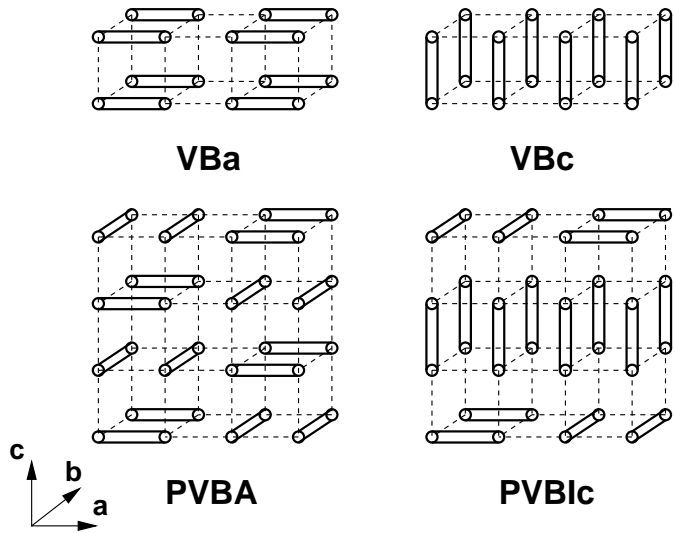}}
\caption{
Spin singlets (double lines) in VB states. The singlets along $a$ axis
(VB$a$) or along $c$ axis (VB$c$) and ordered in either PVBA or PVBI$c$
phase. The axes $\{a,b,c\}$ are shown by arrows.
}
\label{fig:ops}
\end{figure}

Further improvement of the energy might be expected by including
leading quantum fluctuations in the VB (VB$a$ and VB$c$) states. In case
of the VB$a$ phase this would lead to resonance on each plaquette between
the two components composed of spin singlets along the $a$ and $b$ axis,
respectively, but this is hindered by the optimized $z$-like orbitals,
oriented in these states either along the $a$, or along the $b$ axis
(therefore the overlap between these two configurations is severly
reduced). Unlike in the Heisenberg antiferromagnet,
the bonds not occupied by singlets {\it contribute finite energy\/}
due to the orbital terms similar to those in Sec. I, and it turns out to
be better to optimize the orbital energies due to those bonds
depending on whether the bond connects two singlets: either
  ($i$) along a single ($a$, $b$, or $c$) axis, or
 ($ii$) oriented along two different axes with an angle of $\pi/2$, or
($iii$) parallel to each other, with the bond making itself an angle of
        $\pi/2$ to both of them.
As the second type of non-singlet bond is energetically the most
favorable one, more energy is gained if plaquettes occupied by
singlets along $a$ and $b$ axis alternate and form a 3D
superlattice, a plaquette VB alternating (PVBA) state (Fig. 2), giving a
lower energy that the quantum-corrected $A$-AF phase. In contrast, the
energy of the VB$c$ phase is decreased more by the resonance between the
vertical singlets ($\langle ij\rangle\parallel c$). The energy of the
resulting resonating VB$c$ (RVB$c$) state could be obtained using the
Bethe ansatz result for the 1D Heisenberg antiferromagnet,
and adding the orbital energies due to the bonds
$\langle ij\rangle\parallel ab$.

Furthermore, a PVB interlayered phase along the $c$ axis (PVBI$c$), composed
of single planes of the PVBA phase interlayered with two planes of VB$c$
vertical singlets (Fig. 2), is more stable in the crossover regime between
the RVB$c$ and PVBA phases for $varepsilon_z<0$. The orbital
energies $\propto J$ in the bonds which connect the singlets
on the bonds along the $c$ axis
with those lying in the $ab$ planes are then optimized, while the
energies $\propto \varepsilon_z$ are gained in the double layers of VB$c$
phase. For $varepsilon_z>0$, a rotated interlayered phase
(PVBI$a$) is more stable before the PVBA takes over. Altogether, one
finds \cite{Fei97} that an exotic spin-orbital liquid state,
represented here by RVB$c$, PVBI$c$, PVBI$a$ and PVBA phases,
is favored in an extended regime of parameters, in analogy with
a 2D 1/5-depleted lattice \cite{Ued96}, but the present instability is
stronger and happens in a 3D model.

Finally, we remark that the present $d^9$ model (\ref{som9}) applies
also to $d^7$ low-spin Ni$^{3+}$ ions in LiNiO$_2$ via a particle-hole
transformation. However, the situation is rather complex there and
also other models were proposed. First, it was argued that
a model based on symmetry arguments \cite{Ver04}, characterized by a
large number of low-lying singlets associated to dimer coverings of
the triangular lattice, could explain the properties of LiNiO$_2$.
Second, it was shown that charge-transfer terms contribute to the
superexchange and they change the balance between different terms in
the Hamiltonian, making the orbital interactions stronger than the spin
ones \cite{Rei05}. In any case, interplane JT interplane coupling seems
to be too weak in LiNiO$_2$ to stabilize the orbital long-range order.
The observed difference in the physical properties between NaNiO$_2$
and LiNiO$_2$ remains one of the puzzling phenomena in the field
\cite{Rei05} and awaits a future study which has to include all the
above aspects.

\section{Entanglement in spin-orbital models}
\label{sec:enta}

Future developments in the theory to understand better the nature of
ordered and disordered states in spin-orbital systems have to take into
account possible entanglement between both spin and orbital degrees of
freedom on the bonds, as for instance in the states of Fig. 2. It is a
common wisdom that the magnetism of correlated Mott insulators can
be understood by means of the Goodenough-Kanamori rules. They originate
from the mean-field (MF) picture and predict that the superexchange
interaction between two magnetic ions with degenerate orbitals is
strongly AF because of the Pauli principle if the overlap between the
occupied orbitals is large, whereas it is weakly FM when the overlap is
weak or virtually disappears.
This means that spin order and orbital order are complementary ---
      ferro orbital (FO) order supports AF order, while
                     AO  order supports FM spin order.
Indeed, these celebrated rules are well followed in the manganites
\cite{Wei04,Elb05},
where the AO is robust in the FM $ab$ planes, while the orbitals obey
the FO order along the $c$ axis, supporting the AF coupling and leading
to the $A$-AF phase. In fact, one may use here the MF decoupling of the
spin and orbital operators, which is sufficient to explain both the
magnetic and optical properties of LaMnO$_3$ \cite{Ole05}.
The orbital order (or liquid) state determines as well whether the
intersite spin correlations are AF or FM in monolayer \cite{Dag04} and
in bilayer \cite{Dag06} manganites, while in the perovskite
La$_{1-x}$Sr$_x$MnO$_3$ systems the magnons in the FM metallic phase
are well explained by the orbital liquid state \cite{Ole02}.

As a prominent example of the Goodenough-Kanamori complementarity we
would like to mention also the AF phases realized in YVO$_3$
\cite{Ulr03}, which are the subject of intense research in recent years.
While $xy$ orbitals of V$^{3+}$ ions are occupied due to GdFeO$_3$
distortions, the second electron of the $d^2$ ionic configuration may
occupy either $yz$ or $zx$ orbital. Depending on whether these orbitals
follow FO or AO order along the $c$ axis, the magnetic correlations are
there either AF or FM, explaining the origin of the two observed types
of AF order shown in Fig. 3:
($i$) the $C$-AF phase, and
($ii$) the $G$-AF phase (with AF spin order along all three cubic
directions).

%%%%%%%%%%%%%%%%%%%%%%%%%%%%%%%%%%%%%%%%%%%%%%%%%%%%%%%%
%%                             Figure 3
%%%%%%%%%%%%%%%%%%%%%%%%%%%%%%%%%%%%%%%%%%%%%%%%%%%%%%%%
\begin{figure}[t!]
  \epsfysize=5.5cm
  \centerline{\epsfbox{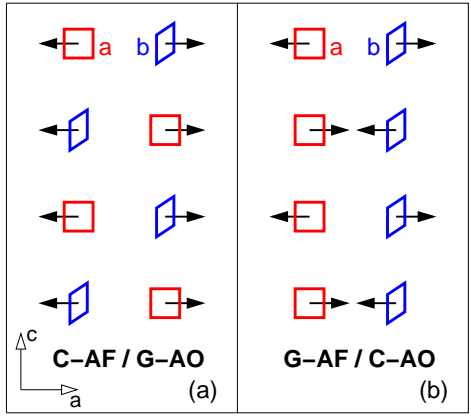}}
\caption{
Goodenough-Kanamori rules on the example of two AF phases observed in
the $R$VO$_3$ perovskites, with complementary magnetic and orbital
order in the $ac$ plane:
(a) $C$-AF spin order accompanied by $G$-AO order; and
(b) $G$-AF spin order accompanied by $C$-AO order.
Arrows indicate spin order, while squares stand for two active
$t_{2g}$ orbitals, $a$ and $b$.
Both spins and orbitals alternate along the $b$ axis (not shown).
}
\label{fig:claso}
\end{figure}

One may verify the Goodenough-Kanamori rules by evaluating intersite
spin and orbital correlations,
\begin{equation}
\label{sij}
S_{ij}=\langle{\vec S}_i\cdot{\vec S}_j\rangle/(2S)^2\,, \hskip 2cm
T_{ij}=\langle{\vec T}_i\cdot{\vec T}_j\rangle\,,
\end{equation}
and comparing them with each other. In addition,
spin-orbital entanglement can be measured by the composite correlation
function defined for a bond $\langle ij\rangle$ \cite{Ole06},
\begin{equation}
\label{cij}
C_{ij}=
\left\{\big\langle({\vec S}_i\cdot{\vec S}_j)
                ({\vec T}_i\cdot{\vec T}_j)\big\rangle
    -\big\langle {\vec S}_i\cdot{\vec S}_j \big\rangle
     \big\langle {\vec T}_i\cdot{\vec T}_j \big\rangle\right\}/(2S)^2\,.
\end{equation}
When $C_{ij}=0$, the spin and orbital operators are disentangled and
their MF decoupling can be applied, while if $C_{ij}<0$ ---
spin and orbital operators are entangled. Consider again the structure
of spin-orbital superexchange models (\ref{som}) derived for $t_{2g}$
systems from charge excitations between
transition metal ions in configurations:
$d^1$ (Ti$^{3+}$ ions in cubic titanates) with $S=1/2$ \cite{Kha00},
and $d^2$ (V$^{3+}$ ions in cubic vanadates) with $S=1$ \cite{Kha01},
for more details about the structure of ${\cal H}_J$ in both $d^1$ and
$d^1$ model see for instance Ref. \cite{Ole05}).

To capture the essence of spin-orbital entanglement it is sufficient to
solve $t_{2g}$ models for $d^1$ and $d^2$ configurations on
four-site chains oriented along the $c$ axis using periodic
boundary conditions. One finds entangled spin-orbital states in both
models, and nontrivial spin-orbital dynamics strongly influences both
intersite spin and orbital intersite correlations (\ref{sij}).
In the $d^1$ (titanate) case one recovers the SU(4) model with
$S_{ij}=T_{ij}=C_{ij}=-0.25$ at $\eta=0$ [Fig. 4(a)].
At finite $\eta$ the SU(4) degeneracy of all intersite correlations is
removed and $T_{ij}<C_{ij}<S_{ij}<0$,
as long as the spin singlet ($S=0$) ground state persists,
i.e., for $\eta< 0.21$. In this regime the Goodenough-Kanamori rule
with complementary correlations (implying different signs of $S_{ij}$
and $T_{ij}$), is violated. Instead, by analyzing the values of the
intersite correlations one finds that the ground state wave function
for each bond $\langle ij\rangle$ is close to a total spin-orbital
singlet, involving a linear combination of
(spin singlet/orbital triplet) and
(spin triplet/orbital singlet) states.
The vanadate $d^2$ model behaves in a similar way, with all three
$S_{ij}$, $T_{ij}$ and $C_{ij}$ correlations being negative in the
spin-singlet ($S=0$) regime of fluctuating $yz$ and $zx$ orbitals
obtained for $\eta< 0.07$ [Fig. 4(b)]. Therefore, the
composite spin-orbital correlations are here finite ($C_{ij}<0$), spin
and orbital variables are {\it entangled\/}, and the MF factorization
of the ground state fails.

%%%%%%%%%%%%%%%%%%%%%%%%%%%%%%%%%%%%%%%%%%%%%%%%%%%%%%%%%%%%
%%                      figure 4
%%%%%%%%%%%%%%%%%%%%%%%%%%%%%%%%%%%%%%%%%%%%%%%%%%%%%%%%%%%%
\begin{figure}[t!]
  \epsfysize=7cm
  \centerline{\epsfbox{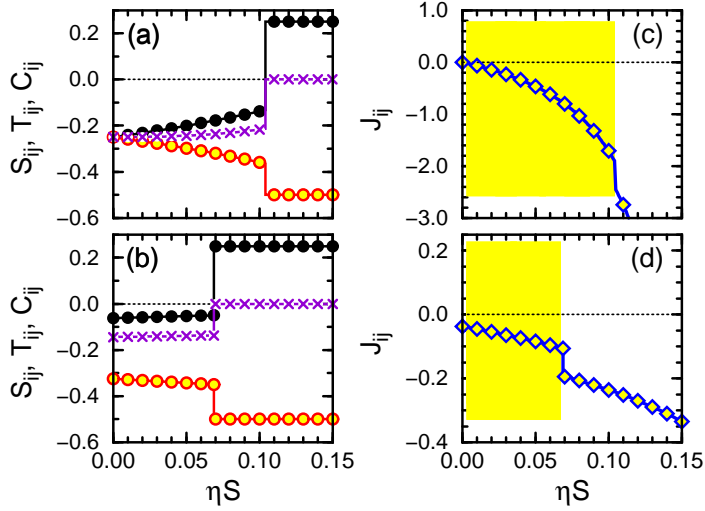}}
\caption{
Left --- intersite
spin    $S_{ij}=\langle{\vec S}_i\!\cdot\!{\vec S}_j\rangle$
(filled circles),
orbital $T_{ij}=\langle{\vec T}_i\!\cdot\!{\vec T}_j\rangle$
(empty circles),
and composite spin-orbital $C_{ij}$ (\ref{cij}) (crosses)
correlations for increasing Hund's exchange $\eta S$ along $c$ axis;
--- right --- the corresponding spin exchange constants
$J_{ij}$ (\ref{cij}),
as obtained for $N=4$ site spin-orbital chain (\ref{som}) with PBC for:
(a),(c) $d^1$ (titanate, $S=1/2$) model, and
(b),(d) $d^2$ (vanadate, $S=1$)   model.
In the shaded areas of (c) and (d) the spin correlations $S_{ij}<0$ do
not follow the sign of the exchange constant $J_{ij}<0$, and the classical
Goodenough-Kanamori rule is violated.
}
\label{fig:enta}
\end{figure}

To provide further evidence that the Goodenough-Kanamori rules do not
apply to $t_{2g}$ systems in the regime of small $\eta$, we compare
spin exchange constants $J_{ij}$ (\ref{Jij}) with the actual values of
intersite spin correlations $S_{ij}$ (\ref{sij}). One finds that
exchange interaction is formally FM ($J_{ij}<0$) in the
orbital-disordered phase at low values of $\eta$ [Figs.
\ref{fig:enta}(c) and \ref{fig:enta}(d)] is in fact accompanied by AF
spin correlations ($S_{ij}<0$), so $J_{ij}S_{ij}>0$ and the ground
state energy calculated in the MF theory is {\it enhanced\/}
\cite{Ole06}. In contrast, similar analysis (not shown) performed for
spin-orbital model (\ref{som9}), derived for $d^9$ ions with $e_g$
orbital degrees of freedom, gave
$J_{ij}S_{ij}<0$, so here spin correlations follow the sign of the
exchange constant \cite{Ole06}. This remarkable difference between
$t_{2g}$ and $e_g$ systems originates from composite spin-orbital
fluctuations, which are responsible for the {\it `dynamical'\/} nature
of exchange constants in the former case. They exhibit large
fluctuations around the average value, measured by $\delta J=
\{\langle ({\hat{\cal J}}_{ij}^{(\gamma)})^2\rangle-J_{ij}^2\}^{1/2}$.
As an illustrative example, we give here the values found in the
$d^1$ and $d^2$ model at $\eta=0$.
While the average spin exchange constant is small in both cases
($J_{ij}\simeq 0$ for $d^1$, $J_{ij}\simeq -0.04$ for $d^2$),
$\hat{\cal J}_{ij}^{(\gamma)}$ fluctuates widely over both positive
and negative values. In the $d^1$ case the fluctuations between
($S=0$/$T=1$) and ($S=1$/$T=0$) bond wave functions are so large that
$\delta J=1$ ! They survive even quite far from the high-symmetry SU(4)
point (at $\eta>0.1$), and stabilize spin-orbital singlet phase in
a broad regime of $\eta$. Also in the $d^2$ model the orbital bond
correlations change dynamically from singlet to triplet,
resulting in $\delta J>|J_{ij}|$, with
$\delta J=\frac{1}{4}\{1-(2T_{ij}+\frac{1}{2})^2\}^{1/2}\simeq 0.25$,
while these fluctuations are small for $d^9$ model (\ref{som9})
with $e_g$ orbitals.

We emphasize that composite spin-orbital fluctuations which occur in
spin-orbital entangled states for realistic parameters determine the
magnetic and optical properties of titanates and vanadates. For
instance, such composite spin-orbital fluctuations are responsible for
the temperature dependence of the optical spectral weights in LaVO$_3$
\cite{Kha04} and trigger spin-orbital dimerization in the $C$-AF phase
of YVO$_3$ in the intermediate temperature regime (see below)
\cite{Hor03}. Remarkably, the observed dimerization in the magnetic
excitations may be sees as a signature of entanglement in excited states
which become relevant at finite temperature. It is activated by thermal
fluctuations in the spin chain \cite{Sir08}, which couple to
dimerized correlations in the orbital sector.

\section{Phase transitions in the $R$VO$_3$ perovskites}
\label{sec:pha}

Recent progress in experimental studies of transition metal oxides
provided exceptionally detailed information on the phase diagrams of
the $R$MnO$_3$ manganites \cite{Goo06} and the $R$VO$_3$ vanadates
\cite{Miy03} (where $R$=La,Pr,$\cdots$,Lu). In the manganites the
orbital order appears first when the temperature is lowered
(at $T_{\rm OO}\sim 800$ K), and its onset is accompanied by a lattice
distortion. The magnetic order follows at much lower temperature
$T_{\rm N}\sim 140$ K. This separation of the spin and orbital energy
scales follows also from the JT distortions which contribute to the
orbital interactions and enhance the value of $T_{\rm OO}$ \cite{Fei99}.
Recent experiments showed that
the orbital transition temperature $T_{\rm OO}$ is enhanced when the
ionic radius $r_R$ of the $R^{3+}$ ions decreases along the $R$MnO$_3$
perovskites, while $T_N$ is drastically reduced, resulting
in the change of magnetic order from the $A$-AF to the $E$-AF phase
\cite{Goo06}. This behavior could not be understood until now and is
one of the challenges for the theory.

\begin{figure}[t!]
  \centerline{\epsfysize=5.1cm\epsfbox{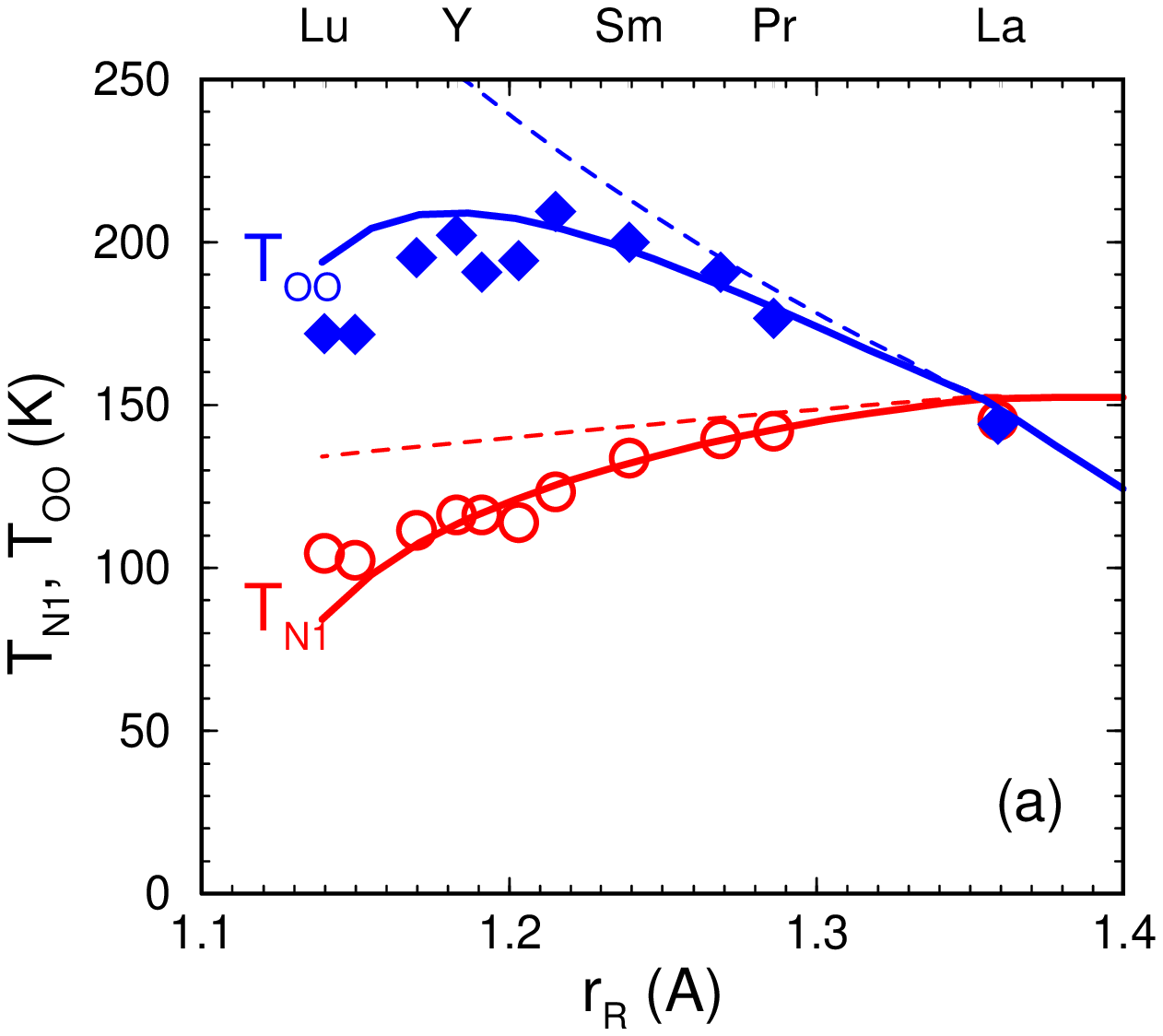}\hskip .5cm
  \epsfysize=5cm\epsfbox{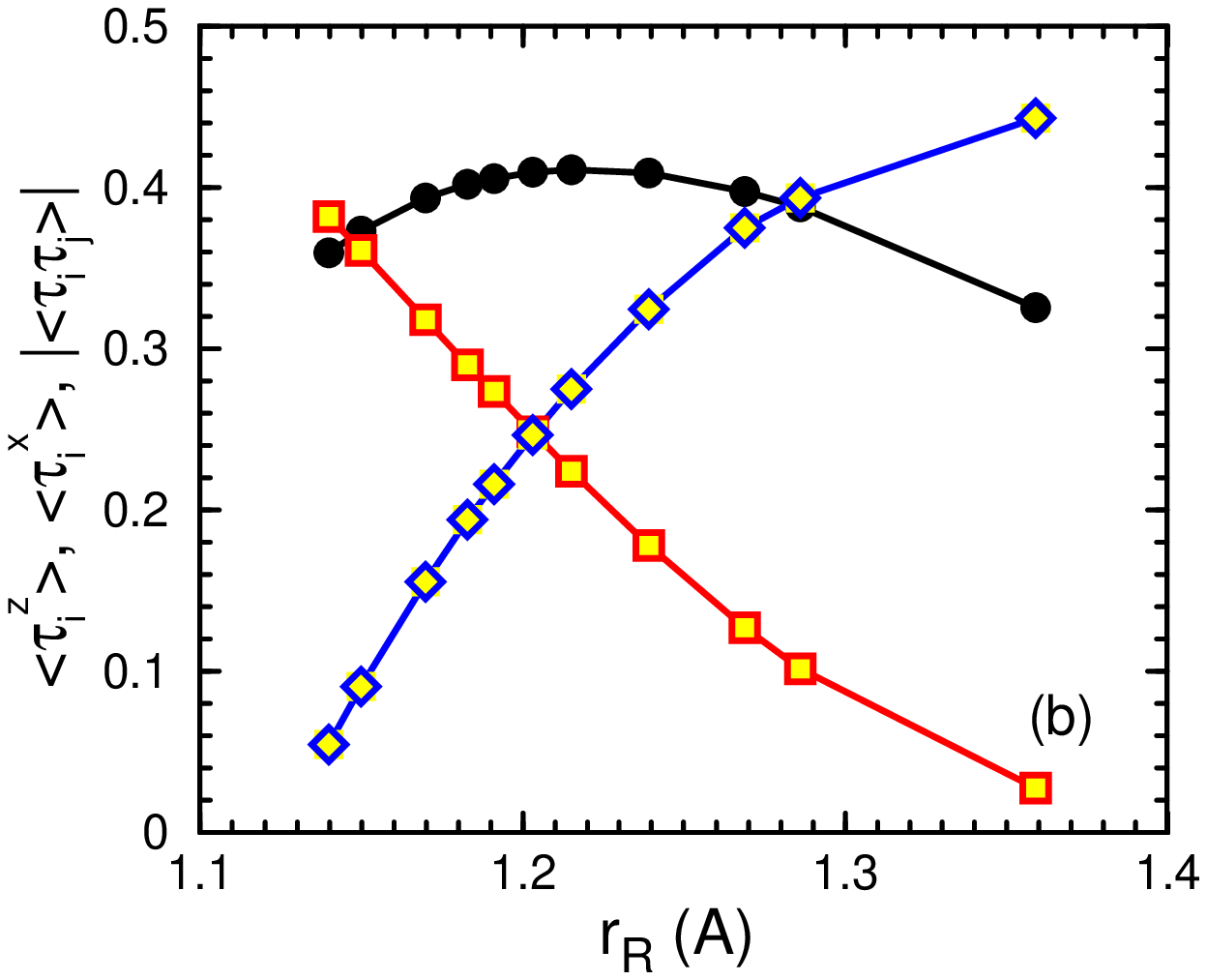}}
\caption{
Theoretical description of orbital and magnetic phase transition in
the $R$VO$_3$ perovskites obtained using model (\ref{som2}),
for varying ionic radius $r_R$ of $R$ ions:
(a) transition temperatures --- $T_{\rm OO}$ for the onset of orbital
and $T_{N1}$ for the onset of the $C$-AF order, as obtained in
Ref. \cite{Hor08} (solid lines), and compared with the experimental
data of Ref. \cite{Miy03} (filled diamonds and open circles);
(b) evolution of orbital order parameter $\langle\tau_i^z\rangle_G$
(filled circles), transverse orbital polarization
$\langle\tau_i^x\rangle$ (squares), and orbital intersite correlations
$|\langle\vec{\tau}_i\cdot\vec{\tau}_j\rangle|$ (diamonds) along
$c$ axis at $T=0$.
Parameters: $J=200$ K, $v_z=17$, $v_{ab}=22$, $v_{g}=740$.
}
\label{fig:rvo}
\end{figure}

The experimental situation in the cubic vanadates is different and even
more complex. On the one hand, the magnetic order in YVO$_3$ is $G$-AF
[Fig. \ref{fig:rvo}(b)] at low temperature, and changes at the first
order magnetic transition at $T_{N2}=77$ K to the $C$-AF structure,
which remains stable up to $T_{N1}\simeq 116$ K. The magnetic transition
at $T_{N2}$ is particularly surprising as the staggered moments change
their direction from approximately parallel to the $c$ axis in the
$G$-AF phase to lying almost within the $ab$ planes in the $C$-AF phase,
with some small alternating $G$-AF component \cite{Ren00}. Although
it was argued that the entropy due to magnetic and orbital excitations is
higher in the $C$-AF phase \cite{Kha01,Ole07}, the magnetization reversal
at the lower magnetic transition remains misterious, and this phenomenon
is still puzzling and far from being completely understood. On the other
hand, only the $C$--AF order develops in LaVO$_3$ below a somewhat higher
$T_{N1}\simeq 143$ K, and is almost immediately followed by a weak
structural transition stabilizing the weak $G$--AO order at
$T_{\rm OO}\simeq 141$ K \cite{Miy03,Miy06} [Fig. \ref{fig:rvo}(a)].
Remarkably, the magnetic order parameter in the $C$-AF phase of LaVO$_3$
is strongly reduced to $\simeq 1.3\mu_B$, much below the reduction
expected from quantum fluctuations in the $C$-AF phase (being only
6\% for $S=1$ spins) \cite{Rac02} --- also this reduction of the
measured magnetization could not be explained so far.

Experimental studies have shown that the $C$-AF order is common to the
entire family of the $R$VO$_3$ vanadates, and in general
$T_{N1}<T_{\rm OO}$, except for LaVO$_3$ with
$T_{N1}\simeq T_{\rm OO}$ \cite{Miy03,Miy06}. When the ionic radius
$r_R$ decreases, the N\'eel temperature $T_{N1}$ also decreases, while
orbital transition temperature $T_{\rm OO}$ increases, passes through a
maximum close to YVO$_3$, and decreases again towards LuVO$_3$.
Knowing that orbital quantum fluctuations and
spin-orbital entanglement play so important role in the perovskite
vanadates, it is of interest to ask whether the spin-orbital model
of the form (\ref{som}) introduced for the perovskite vanadates
\cite{Kha01,Hor03} is able to describe this variation of $T_{\rm OO}$
and $T_{N1}$ with decreasing radius $r_R$ of $R$ ions in $R$VO$_3$
\cite{Miy03}. It is clear that the nonmonotonic dependence of
$T_{\rm OO}$ on $r_R$ cannot be reproduced just by the superexchange,
as it requires two mechanisms which oppose each other. In fact, there
is even no reason to assume that the superexchange constant $J$ should
depend on $r_R$, as the distances between $V^{3+}$ ions are quite close
to each other in different compounds \cite{Ree06,Sag06,Sag07}, so
neither $t$ nor $U$ is expected to change significantly.

As in the $R$MnO$_3$ manganites \cite{Goo06}, one expects that the JT
distortions should increase when the ionic radius $r_R$ decreases, and
one could argue that this would induce the increase of $T_{\rm OO}$. In
order to unravel the physical mechanism responsible for the decrease of
$T_{\rm OO}$ from YVO$_3$ to LuVO$_3$ one has to analyze in more detail
the evolution of GdFeO$_3$ distortions with decreasing $r_R$
\cite{Hor08}. Such distortions are common for the perovskites
\cite{Pav05}, and may be described by two subsequent rotations of VO$_6$
octahedra:
($i$) by an angle $\vartheta$ around the $b$ axis, and
($ii$) by an angle $\varphi$ around the $c$ axis.
Increasing angle $\vartheta$ causes a decrease of
V--O--V bond angles along the $c$ direction, being $\pi-2\vartheta$,
and leads to an orthorhombic lattice distortion $u=(b-a)/a$, where
$a$ and $b$ are the lattice parameters of the $Pbnm$ structure of
$R$VO$_3$. The structural data for the perovskite $R$VO$_3$
vanadates \cite{Ree06,Sag06,Sag07} give the following empirical
relation between the ionic radius $r_R$ and the angle $\vartheta$:
\begin{equation}
\label{r}
r_R=r_0-\alpha\sin^2\vartheta\,,
\end{equation}
where $r_0=1.5$ \AA{} and $\alpha=0.95$ \AA{}. This allows one to
use the angle $\vartheta$ to parametrize the dependence of the
microscopic parameters and the transition temperatures
$T_{\rm OO}$ and $T_{N1}$ on $r_R$.

The spin-orbital model introduced in Ref. \cite{Hor08} to describe
the phase diagram of $R$VO$_3$ includes:
($i$) the superexchange $\propto J$ between $V^{3+}$ ions in the $d^2$
configuration with $S=1$ spins \cite{Kha01},
($ii$) the crystal-field splitting $\propto E_z$ between $yz$ and
$zx$ orbitals,
($iii$) intersite orbital interactions $\propto V_{ab},V_c$ (which
originate from the coupling to the lattice), and
($iv$) orbital-lattice term which induces orbital polarization when
the distortion $u$ inreases.
The Hamiltonian consists thus of several terms \cite{Hor08},
\begin{eqnarray}
\label{som2}
{\cal H}\!\!\!&=&\!\!\!J\!\!\sum_{\langle ij\rangle\parallel\gamma}\!
\left\{\!\Big({\vec S}_i\!\cdot\!{\vec S}_j\!+\!S^2\Big){{\cal
J}}_{ij}^{(\gamma)}
+ {{\cal K}}_{ij}^{(\gamma)}\!\right\}
+E_z(\vartheta)\!\sum_i\!e^{i{\vec R}_i{\vec Q}}\tau_i^z
- V_{c}(\vartheta)\!\!\sum_{\langle ij\rangle\parallel c}\!\!
\tau_i^z\tau_j^z
\nonumber \\
&+&V_{ab}(\vartheta)\!\sum_{\langle ij\rangle\parallel ab}\tau_i^z\tau_j^z
-gu\sum_i\tau_i^x+\frac12 N K(u-u_0(\vartheta))^2\,.
\end{eqnarray}
with $\gamma=a,b,c$ labels the cubic axes.
The orbital operators take the form:
\begin{eqnarray}
\label{orbj}
{\hat J}_{ij}^{(\gamma)}&=&
\frac{1}{2}\left\{(1+2\eta r_1)
\left({\vec\tau}_i\cdot {\vec\tau}_j
     +\frac{1}{4}n_i^{}n_j^{}\right)\right.          \nonumber \\
&-&\! \left.\eta r_3
    \left({\vec\tau}_i\times{\vec\tau}_j+\frac{1}{4}n_i^{}n_j^{}\right)
-\frac{1}{2}\eta r_1(n_i+n_j)\right\}^{(\gamma)},             \\
\label{orbk}
{\hat K}_{ij}^{(\gamma)}&=&
\left\{\eta r_1
\left({\vec\tau}_i\cdot {\vec\tau}_j+\frac{1}{4}n_i^{}n_j^{}\right)
 +\eta r_3\left({\vec\tau}_i\times {\vec\tau}_j
             +\frac{1}{4}n_i^{}n_j^{}\right)\right.  \nonumber \\
&-&\!\left.
   \frac{1}{4}(1+\eta r_1)(n_i+n_j)\right\}^{(\gamma)},
\end{eqnarray}
and have a rich structure which originates from the projections of the
$d^3_i$ excited states on the respective eigenstates of V$^{2+}$ ion.
They arise due to the
$d_i^2d_j^2\rightleftharpoons d_i^3d_j^1$ charge excitations,
leading either to high-spin or to low-spin $d^3_i$ configurations,
so Hund's exchange splittings in the
multiplet structure of a V$^{2+}$ ions enter via the coefficients
$r_1=1/(1-3\eta)$ and $r_3=1/(1+2\eta)$ (the lower energy singlet
excitations occur at energy $U$, so the corresponding coefficient is
$r_2=1$). Using the atomic value of Hund's exchange, we have estimated
that $\eta\simeq 0.13$ \cite{Ole05,Ole07}. The orbital operators
${\vec\tau}_i=\{\tau_i^+,\tau_i^-,\tau_i^z\}$ for pseudospin $\tau=1/2$
in Eqs. (\ref{orbj}) and (\ref{orbk}) are defined in the subspace
spanned by two orbital flavors which are active along a given cubic
direction $\gamma=a,b,c$. Unlike in $d^9$ model (\ref{som9}) for $e_g$
orbitals, the leading orbital interactions are here proportional to the
scalar products $({\vec\tau}_i\cdot {\vec\tau}_j)^{(\gamma)}$ of orbital
operators on the bonds, but the structure of local
Coulomb interactions is responsible for additional terms,
\begin{equation}
\left({\vec\tau}_i\times {\vec\tau}_j\right)^{(\gamma)}=
\frac{1}{2}\left(\tau_i^+\tau_j^++\tau_i^-\tau_j^-\right)
+\tau_i^z\tau_j^z\,,
\end{equation}
so the orbital quantum numbers are not conserved.
Finally, the operator $n_i^{(\gamma)}$ stands for the number of active
electrons at site $i$ along the bond $\langle ij\rangle$, for instance
for a bond along the $c$ axis this number is
$n_i^{(c)}=n_{i,yz}^{}+n_{i,zx}^{}$.

Further insight into the electronic configuration on V$^{3+}$ ions in
$R$VO$_3$ may be obtained either by investigating the electronic
structure \cite{And07}, or by explicit calculations using the point
charge model \cite{Hor08}. Both lead to the electronic configurations
$(xy)^1(yz/zx)^1$, i.e., the $xy$ orbitals are occupied at all sites,
while the remaining $\{yz,zx\}$ orbitals represent active orbital
degrees of freedom which contribute to the $t_{2g}$ orbital dynamics,
expressed by the scalar product ${\vec\tau}_i\cdot {\vec\tau}_j$ (only
for $\gamma\equiv c$). This demonstrates an important difference
between the $e_g$ (with one electron or one hole in $e_g$ orbitals at
each site) and a $t_{2g}$ system, such as the $R$VO$_3$ perovskites.
Two active $t_{2g}$ orbitals along this bond open a new possibility ---
if both orbitals are singly occupied, an {\it orbital singlet\/} gives
here FM superexchange, even in the absence of Hund's exchange
(at $\eta=0$) \cite{Kha01}.

Furthermore, the actual electronic configurations realized in the
$R$VO$_3$ systems justify the form of the crystal-field term used in
Eq. (\ref{som2}), with $\tau_i^z=\frac{1}{2}(n_{i,yz}-n_{i,zx})$. The
crystal field alternates in the $ab$ planes, but is uniform along the
$c$ axis. It is thus characterized by the vector ${\vec Q}=(\pi,\pi,0)$
in reciprocal space, and competes with the (weak) $G$-AO order
supporting the observed $C$-AF phase below $T_{N1}$.
The orbital interactions induced by the distortions of the VO$_6$
octahedra and by GdFeO$_3$ distortions of the lattice, $V_{ab}>0$ and
$V_c>0$, also favor the $C$-AO order (like $E_z>0$). Note that $V_c>0$
counteracts the orbital interactions included in the superexchange.
In case of $V_{c}$ one may assume that its dependence on the angle
$\vartheta$ is weak, and a constant $V_{c}(\vartheta)=0.26J$
was chosen in Ref. \cite{Hor08} in order to satisfy the experimental
condition that the $C$-AF and $G$-AO order appears almost
simultaneously in LaVO$_3$, see Fig. \ref{fig:rvo}(a). The
experimental value $T_{N1}^{\rm exp}=143$ K for LaVO$_3$ \cite{Miy03}
was fairly well reproduced in the present model taking $J=200$ K.
The functional dependence of the remaining two
parameters $\{E_z,V_{ab}\}$ on the tilting angle $\vartheta$ was
derived from the point charge model \cite{Hor08}, using the structural
data for $R$VO$_3$ \cite{Ree06,Sag06,Sag07}. In this way the following
two relations were deduced:
\begin{eqnarray}
\label{Ez}
E_z(\vartheta)&=&J\,v_z\,\sin^3\vartheta\cos\vartheta\,, \\
\label{vab}
V_{ab}(\vartheta)&=&J\,v_{ab}\,\sin^3\vartheta\cos\vartheta\,.
\end{eqnarray}

The last two terms in Eq. (\ref{som2}) describe the linear coupling
$\propto g>0$ between active $\{yz,zx\}$ orbitals and the orthorhombic
lattice distortion $u$. The elastic energy which counteracts lattice
distortion $u$ is given the force constant $K$, and
$N$ is the number of $V^{3+}$ ions. The coupling $\propto gu$ acts as
a transverse field in the pseudospin space. While the eigenstates
$\frac{1}{\sqrt{2}}(|yz\rangle\pm|zx\rangle)$ favored by $\tau_i^x$
cannot be realized due to the competition with all the
other terms, increasing lattice
distortion $u$ (increasing angle $\vartheta$) modifies the
orbital order and intersite orbital correlations. At the minimum
one finds an effective coupling,
\begin{equation}
\label{g}
g_{\rm eff}(\vartheta;T)\equiv gu(\vartheta;T)
=gu_0(\vartheta)+\frac{g^2}{K}\langle\tau^x\rangle_T\,,
\end{equation}
with the global distortion $u(\vartheta;T)$ caused by
($i$) a pure lattice term $u_0(\vartheta)$, and
($ii$) an orbital contribution $\propto\langle\tau^x\rangle$ which can
be determined self-consistently within the present approach, see Ref.
\cite{Hor08}.
Both $u_0$ and $\langle\tau^x\rangle$ are expected to increase with
increasing tilting $\vartheta$. This dependence has to be faster than
the ones of Eqs. (\ref{Ez}) and (\ref{vab}), as otherwise no maximum
in the dependence of $T_{\rm OO}$ on $r_R$ would be obtained. Using
this argument a semiempirical relation,
\begin{equation}
\label{geff}
g_{\rm eff}(\vartheta)=J\,v_{g}\,\sin^5\vartheta\cos\vartheta\,,
\end{equation}
was postulated in Ref. \cite{Hor08}. Therefore, magnetic and orbital
correlations described by spin-orbital model (\ref{som}) depend on
three parameters: $\{v_z,v_{ab},v_g\}$.

%%%%%%%%%%%%%%%%%%%%%%%%%%%%%%%%%%%%%%%%%%%%%%%%%%%%%%%%%%%%
%%                      figure 5
%%%%%%%%%%%%%%%%%%%%%%%%%%%%%%%%%%%%%%%%%%%%%%%%%%%%%%%%%%%%

While the above fast dependence on the tilting angle $\vartheta$ of
VO$_6$ octahedra in the $R$VO$_3$ family was introduced in order to
reproduce the experimentally observed dependence of $T_{\rm OO}$ on
$r_R$, see Fig. \ref{fig:rvo}(a), it may
be justified {\it a posteriori\/}. It turns out that
the dependence of $g_{\rm eff}$ on the ionic radius $r_R$
follows the actual lattice distortion $u$ in $R$VO$_3$
measured at $T=0$ (or just above $T_{N1}$) \cite{Hor08}. Also
the orbital polarization $\langle\tau^x\rangle$ is approximately
$\propto\sin^5\vartheta\cos\vartheta$, and follows the same fast
dependence of $g_{\rm eff}(\vartheta)$ for the $R$VO$_3$ perovskites.
These results indicate that the bare coupling parameters $\{g,K\}$ are
nearly constant and independent of $r_R$, which may be treated as a
prediction of the theory to be verified by future experiments.

Similar to the $R$MnO$_3$ manganites \cite{Fei99}, the correct
MF treatment of the orbital and magnetic phase transitions in the
$R$VO$_3$ vanadates requires the coupling between the on-site orbital,
$\langle\tau^z\rangle_G\equiv\frac12|\langle\tau^z_i-\tau^z_j\rangle|$,
and spin, $\langle S_i^z\rangle_C$, order parameters in the $C$-AF
phase, as well as a composite $\langle S_i^z\tau_i^z\rangle$ order
parameter. Unlike in the perovskite manganites
\cite{Goo06}, however, the on-site MF theory does not suffice
for the $R$VO$_3$ compounds as the orbital
singlet correlations $\langle{\vec\tau}_i\cdot{\vec\tau}_j\rangle$
on the bonds $\langle ij\rangle$ along the $c$ axis play an important
role in stabilizing the $C$-AF phase \cite{Kha01,Ole07}. Therefore,
the minimal physically acceptable approach to the present problem is
a self-consistent calculation for a bond $\langle ij\rangle$ along
the $c$ axis, embedded in the MF terms due to its neighbors along all
the cubic axes \cite{Hor08}. This procedure, with properly selected
model parameters, served to reproduce the experimental phase diagram
of Fig. \ref{fig:rvo}(a). One finds that the orbital transition occurs
first in the $R$VO$_3$ vanadates with $r_R<r_{\rm La}$, as observed.
At the same time, the N\'eel temperature $T_{N1}$ decreases with
decreasing $r_R$.

The key feature of the present spin-orbital system which drives the
observed dependence of $T_{\rm OO}$ and $T_{N1}$ on the ionoc radius
$r_R$ is the evolution of intersite orbital correlations
$\langle{\vec\tau}_i\cdot{\vec\tau}_j\rangle$ along the $c$ axis,
and the increasing orbital polarization $\langle\tau^x\rangle$ with
decreasing $r_R$ [Fig. \ref{fig:rvo}(b)]. Indeed, the singlet
correlations are drastically suppressed from LaVO$_3$ towards LuVO$_3$.
While $\langle\tau^x\rangle\simeq 0.03$ is rather weak in LaVO$_3$,
it steadily increases along the $R$VO$_3$ perovskites when $r_R$
decreases, and finally it becomes as important as the orbital order
parameter, $\langle\tau^x\rangle\simeq\langle\tau^z\rangle_G$. Note
that in all the cases the latter order parameter is substantially
reduced from the classical value $\frac12$ by singlet orbital
fluctuations, being $\langle\tau^z\rangle_G\simeq 0.32$ and 0.36 for
LaVO$_3$ and LuVO$_3$, respectively.

The results presented in Ref. \cite{Hor08} demonstrate the remarkable
dependence of both spin and orbital order on the orbital correlations.
First, the increase of orbital intersite interactions due to the JT
term (\ref{vab}), induces steady increase of the orbital temperature
$T_{\rm OO}$ with decreasing $r_R$. When the orbital polarization
$\langle\tau^x\rangle$ becomes large, however, this increase is
suppressed and one reproduces the nonmonotonic dependence of
$T_{\rm OO}$ on $r_R$, with the observed drop of $T_{\rm OO}$
when $r_R$ decreases beyond $r_R\sim 1.18$ \AA{} of YVO$_3$.
Second, the changes in intersite orbital correlations shown in Fig.
\ref{fig:rvo}(b) modify the magnetic exchange constants
$\{J_{ab},J_c\}$ along the bonds parallel to one of cubic directions
in the $ab$ planes or along the $c$ axis (\ref{Jij}), and thus the
value of $T_{N1}$ is reduced with decreasing $r_R$, although the
superexchange energy $J$ does not change. This also means that the
width of the magnon band given at
$T=0$ by $W_{C-{\rm AF}}=4(J_{ab}+|J_c|)$ \cite{Ole07} is reduced by
a factor close to 1.8 from LaVO$_3$ to YVO$_3$, in agreement with
surprisingly low magnon energies observed in the $C$-AF phase of
YVO$_3$ \cite{Ulr03}.

Summarizing, spin-orbital model (\ref{som2}) of Ref. \cite{Hor08}
provides an almost quantitative understanding of the systematic
experimental trends for both orbital and magnetic transitions in the
$R$VO$_3$ perovskites, and is able to reproduce the nonmonotonic
variation of the orbital temperature $T_{\rm OO}$ for decreasing $r_R$.
Hovever, the theoretical description of the magnetic transition to the
$G$-AF phase at $T_{N2}$, which occurs for small $r_R$ \cite{Miy03},
remains to be addressed by future theory. More open issues and future
directions of reasearch in the field of the orbital physics are
shortly indicated in the next section.

\section{Summary and open problems}
\label{sec:sum}

A few representative problems discussed above demonstrate that the
spin-orbital physics is a very rich field, with intrinsically
frustatred interactions and rather exotic ordered or disordered phases,
and with their behavior frequently dominated by quantum fluctuations.
While valuable information about the electronic structure is obtained
from density functional theory \cite{Sol08}, the many-body aspects have
to be studied using models of correlated electrons. Perhaps the most
important feature of orbital or spin-orbital superexchange models is
strong coupling between the orbitals and the lattice, which may help
to stabilize orbital order, as in model (\ref{som2}) introduced for
the $R$VO$_3$ perovskites \cite{Hor08}. In this way, the lattice
distortions may also indirectly influence the onset of magnetic order in
systems with active orbital degrees of freedom. If they are absent and
the lattice is frustrated in addition, a very interesting situation
arises, with strong tendency towards truly exotic quantum states
\cite{Kha05}. Examples of this behavior were considered recently for the
triangular lattice, both for $e_g$ orbitals in LiNiO$_2$ \cite{Rei05}
and $t_{2g}$ orbitals in NaTiO$_2$ \cite{Nor08}. None of these models
could really be solved, but generic tendency towards dimer correlations
with spin singlets on the bonds for particular orbital states could be
shown.

Rapid progress is the field of orbital physics results mainly from
developing new experimental techniques and synthesizing novel materials.
While the experiment is ahead of the theory in most cases, there are
some exceptions. One of them was a theoretical prediction of the energy
and dispersion of orbital excitations \cite{vdB99,vdB01}. Only recently
orbital excitations (orbitons) could be observed in Raman scattering in
the Mott insulators LaTiO$_3$ and YTiO$_3$ \cite{Ulr06}. An interesting
question which arises in this context is carrier propagation in a Mott
insulator with AO order. This problem is very complex and has been
addressed also in the last decade of the previous century \cite{Zaa93}.
Not only spin excitations may provide more final states, as for triplet
excitations in the $t$-$J$ model \cite{Zaa92}, but a carrier may dress
by orbital excitations \cite{vdB00}, and scatter on magnons
\cite{Bal01}. Indeed, the coupling to orbitons increases the effective
mass of a moving hole in $e_g$ systems \cite{vdB00}. The orbital part of
the superexchange is classical $t_{2g}$ systems, but also there weak
quasiparticle dispersion arises from three-site processes
\cite{Dag08} (see also this volume \cite{Woh08}).

Similar to doped manganites \cite{Dag07}, also in doped
$R_{1-x}$(Sr,Ca)$_x$VO$_3$ systems orbital order gradually disappears
\cite{Fuj05}. The composite $C$-AF/$G$-AO order survives, however, is
a broad range of doping, in contrast to La$_{1-x}$Sr$_x$MnO$_3$, where
FM order sets in already at $x\sim 0.10$. It is quite remarkable that
the complementary $G$-AF/$C$-AO type of order is fragile and disappears
in Y$_{1-x}$Ca$_x$VO$_3$ already at $x=0.02$ \cite{Fuj05}. We remark
that the hole doped in $C$-AF/$G$-AO phase are localized in polaronlike
states \cite{Fuj08}, so the purely electronic model such as of Ref.
\cite{Dag08} is too crude to capture both the evolution of the
spin-orbital order in doped vanadates and the gradual decrease of the
energy scale for spin-orbital fluctuations.

As a final remark, we would like to mention very promissing recent
experimental studies of Ni-based superlattices \cite{Cha06}. Recent
theory for LaNiO$_3$/La$M$O$_3$ superlattices (with $M$=Al, Gd, Ti)
\cite{Cha08} predicts that the correlated $e_g$ electrons in the
NiO$_2$ planes develop a planar $x^2-y^2$ orbital order driven by the
reduced dimensionality and further supported by epitaxial strain from
the substrate. This resembles the $x^2-y^2$ orbital polarization in
doped layered manganites \cite{Mac99}. As in all other cases discussed
above, the superexchange interactions which involve the orbital degrees
of freedom play here a crucial role to understand the observed
magnetic order and low-lying excited states.

\section*{Acknowledgments}
It is a great pleasure to thank all my collaborators, in particular
P. Horsch, L.F. Feiner, G. Khaliullin and J. Zaanen,
for numerous insightful discussions which contributed
to my present understanding of the subject.
We acknowledge support by the Foundation for Polish Science
(FNP), and by the Polish Ministry of Science and Higher Education
under Project No.~N202 068 32/1481.


\begin{thebibliography}{}

\bibitem{Ima98} M. Imada, A. Fujimori and Y. Tokura,
                   {\em Rev. Mod. Phys.\/} \textbf{70}, 1039 (1998).

\bibitem{Kug82} K.I. Kugel and D.I. Khomskii,
                   {\em Sov. Phys. Usp.\/} \textbf{25}, 231 (1982).

\bibitem{Cnr78} C. Castellani, C.R. Natoli and J. Ranninger,
                    {\em Phys. Rev. B\/} \textbf{18}, 4945 (1978);
                        \textbf{18}, 4967 (1978);
                        \textbf{18}, 5001 (1978).

\bibitem{Fei99} L.F. Feiner and A.M. Ole\'s,
                   {\em Phys. Rev. B\/} \textbf{59}, 3295 (1999).

\bibitem{Fei97} L.F. Feiner, A.M. Ole\'s and J. Zaanen,
                   {\em Phys. Rev. Lett.\/} \textbf{78}, 2799 (1997).

\bibitem{Ole06} A.M. Ole\'s, P. Horsch, L.F. Feiner and G. Khaliullin,
                   {\em Phys. Rev. Lett.\/} \textbf{96}, 147205 (2006).

\bibitem{NJP}   B. Keimer and A.M. Ole\'s, {\it Focus on Orbital Physics\/},
                {\em New J. Phys.\/} \textbf{6}, E05 (2004).

\bibitem{Ole05} A.M. Ole\'s, P. Horsch, G. Khaliullin and L.F. Feiner,
                   {\em Phys. Rev. B\/} \textbf{72}, 214431 (2005).

\bibitem{Cha77} K.A. Chao, J. Spa\l{}ek and A.M. Ole\'s,
                   {\em J. Phys. C\/} \textbf{10}, L271 (1977).

\bibitem{vdB99} J. van~den Brink, F. Mack, P. Horsch and A.M. Ole\'s,
                   {\em Phys. Rev. B\/} \textbf{59}, 6795 (1999).

\bibitem{Ole00} A.M. Ole\'s, L.F. Feiner and J. Zaanen,
                   {\em Phys. Rev. B\/} \textbf{61}, 6257 (2000).

\bibitem{vdB04} J. van den Brink,
                   {\em New J. Phys.\/} \textbf{6}, 201 (2004).

\bibitem{Fei05} L.F. Feiner and A.M. Ole\'s,
                   {\em Phys. Rev. B\/} \textbf{71}, 144422 (2005).

\bibitem{Kho03} D.I. Khomskii and M.V. Mostovoy,
           {\em J. Phys. A\/} \textbf{36}, 9197 (2003).

\bibitem{Mil05} J. Dorier, F. Becca, and F. Mila,
                   {\em Phys. Rev. B\/} \textbf{72}, 024448 (2005).

\bibitem{Wen08} S. Wenzel and W. Janke,
                   {\em Phys. Rev. B\/} \textbf{78}, 064402 (2008).

\bibitem{Lon80} L. Longa and A.M. Ole\'s,
                   {\em J. Phys. A\/} \textbf{13}, 1031 (1980).

\bibitem{Brz07} W. Brzezicki, J. Dziarmaga and A.M. Ole\'s,
                   {\em Phys. Rev. B\/} \textbf{75}, 134415 (2007).

\bibitem{Perk}  J.H.H. Perk {\it et al.},
                   {\em Physica A} \textbf{123}, 1 (1984).

\bibitem{Brz08} W. Brzezicki and A.M. Ole\'s,
                   {\em Acta Phys. Polon. A\/} \textbf{115}, this issue (2008).

\bibitem{Kha04} G. Khaliullin, P. Horsch and A.M. Ole\'s,
                   {\em Phys. Rev. B\/} \textbf{70}, 195103 (2004).

\bibitem{Fei98} L.F. Feiner, A.M. Ole\'s and J. Zaanen,
                   {\em J. Phys.: Condens. Matter\/} \textbf{10}, L555 (1997).

\bibitem{Cha88} P. Chandra and B. Dou\c{c}ot,
                   {\em Phys. Rev. B\/} \textbf{38}, 9335 (1988).

\bibitem{Kha98} G. Khaliullin and V. Oudovenko,
                   {\em Phys. Rev. B\/} \textbf{56}, 14243(R) (1997).

\bibitem{Ued96} K. Ueda, H. Kontani, M. Sigrist and P.A. Lee,
                   {\em Phys. Rev. Lett.\/} \textbf{76}, 1932 (1996).

\bibitem{Ver04} F. Vernay, K. Penc, P. Fazekas and F. Mila,
                   {\em Phys. Rev. B\/} \textbf{70}, 014428 (2004).

\bibitem{Rei05} A.J.W. Reitsma, L.F. Feiner and A.M. Ole\'s,
                   {\em New J. Phys.\/} \textbf{7}, 121 (2005).

\bibitem{Wei04} A. Wei\ss{}e and H. Fehske,
                   {\em New J. Phys.\/} \textbf{6}, 158 (2004).

\bibitem{Elb05} E. Dagotto,
                   {\em New J. Phys.\/} \textbf{7}, 67 (2005).

\bibitem{Dag04} M. Daghofer, W. von der Linden and A.M. Ole\'s,
                   {\em Phys. Rev. B\/} \textbf{70}, 184430 (2004).

\bibitem{Dag06} M. Daghofer, A.M. Ole\'s, D.M. Neuber and W. von der Linden,
                   {\em Phys. Rev. B\/} \textbf{73}, 104451 (2006).

\bibitem{Ole02} A.M. Ole\'s and L.F. Feiner,
                   {\em Phys. Rev. B\/} \textbf{65}, 052414 (2002).

\bibitem{Ulr03} C. Ulrich, G. Khaliullin, J. Sirker, M. Reehuis,
                   M. Ohl, S. Miyasaka, Y. Tokura and B. Keimer,
                   {\em Phys. Rev. Lett.\/} \textbf{91}, 257202 (2003).

\bibitem{Kha00} G. Khaliullin and S. Maekawa,
                   {\em Phys. Rev. Lett.\/} \textbf{85}, 3950 (2000).

\bibitem{Kha01} G. Khaliullin, P. Horsch and A.M. Ole\'s,
                   {\em Phys. Rev. Lett.\/} \textbf{86}, 3879 (2001).

\bibitem{Hor03} P. Horsch, G. Khaliullin and A.M. Ole\'s,
                   {\em Phys. Rev. Lett.\/} \textbf{91}, 257203 (2003).

\bibitem{Sir08} J. Sirker, A. Herzog, A.M. Ole\'s and P. Horsch,
                   {\em Phys. Rev. Lett.\/} \textbf{101}, 157204 (2008).

\bibitem{Goo06} J.-S. Zhou and J.B. Goodenough,
                   {\em Phys. Rev. Lett.\/} \textbf{96}, 247202 (2006).

\bibitem{Miy03} S. Miyasaka, Y. Okimoto, M. Iwama and Y. Tokura,
                   {\em Phys. Rev. B\/} \textbf{68}, 100406 (2003).

\bibitem{Ren00} Y. Ren, T.T.M. Palstra, D.I. Khomskii,
                   A.A. Nugroho, A.A. Menovsky and G.A. Sawatzky,
                   {\em Phys. Rev. B\/} \textbf{62}, 6577 (2000).

\bibitem{Ole07} A.M. Ole\'s, P. Horsch and G. Khaliullin,
                   {\em Phys. Rev. B\/} \textbf{75}, 184434 (2007).

\bibitem{Miy06} S. Miyasaka, J. Fujioka, M. Iwama, Y. Okimoto
                   and Y. Tokura,
                   {\em Phys. Rev. B\/} \textbf{73}, 224436 (2006).

\bibitem{Rac02} M. Raczkowski and A.M. Ole\'s,
                   {\em Phys. Rev. B\/} \textbf{66}, 094431 (2002).

\bibitem{Ree06} M. Reehuis, C. Ulrich, P. Pattison, B. Ouladdiaf,
                   M.C. Rheinstädter, M. Ohl, L.P. Regnault, M. Miyasaka,
                   Y. Tokura and B. Keimer,
                   {\em Phys. Rev. B\/} \textbf{73}, 094440 (2006).

\bibitem{Sag06} M.H. Sage, G.R. Blake, G.J. Nieuwenhuys and T.T.M. Palstra,
                   {\em Phys. Rev. Lett.\/} \textbf{96}, 036401 (2006).

\bibitem{Sag07} M.H. Sage, G.R. Blake, C. Marquina and T.T.M. Palstra,
                   {\em Phys. Rev. B\/} \textbf{96}, 195102 (2007).

\bibitem{Hor08} P. Horsch, A.M. Ole\'s, L.F. Feiner and G. Khaliullin,
                   {\em Phys. Rev. Lett.\/} \textbf{100}, 167205 (2008).

\bibitem{Pav05} E. Pavarini, Y. Yamasaki, J. Nuss and O.K. Andersen,
                   {\em New J. Phys.\/} \textbf{7}, 188 (2005).

\bibitem{And07} M. De Raychaudhury, E. Pavarini and O.K. Andersen,
                   {\em Phys. Rev. Lett.\/} \textbf{99}, 126402 (2007).

\bibitem{Sol08} I.V. Solovyev,
                   {\em J. Phys.: Condens. Matter\/} \textbf{20}, 293201 (2008).

\bibitem{Kha05} G. Khaliullin,
                   {\em Prog. Theor. Phys. Suppl.\/} \textbf{160}, 155 (2005).

\bibitem{Nor08} B. Normand and A.M. Ole\'s,
                   {\em Phys. Rev. B\/} \textbf{78}, 094427 (2008).

\bibitem{vdB01} J. van~den Brink,
                   {\em Phys. Rev. Lett.\/} \textbf{87}, 217202 (2001).

\bibitem{Ulr06} C. Ulrich, A. G\"ossling, M. Gr\"uninger, M. Guennou,
                   H. Roth, M. Cwik, T. Lorenz, G. Khaliullin
                   and B. Keimer,
                   {\em Phys. Rev. Lett.\/} \textbf{97}, 157401 (2006).

\bibitem{Zaa93} J. Zaanen and A.M. Ole\'s,
                   {\em Phys. Rev. B\/} \textbf{48}, 7197 (1993).

\bibitem{Zaa92} J. Zaanen, A.M. Ole\'s and P. Horsch,
                   {\em Phys. Rev. B\/} \textbf{46}, 5798 (1992).

\bibitem{vdB00} J. van~den Brink, P. Horsch and A.M. Ole\'s,
                   {\em Phys. Rev. Lett.\/} \textbf{85}, 5174 (2000).

\bibitem{Bal01} J. Ba\l{}a, G.A. Sawatzky, A.M. Ole\'s and A. Macridin,
                   {\em Phys. Rev. Lett.\/} \textbf{87}, 067204 (2001).

\bibitem{Dag08} M. Daghofer, K. Wohlfeld, A.M. Ole\'s, E. Arrigoni
                   and P. Horsch,
                   {\em Phys. Rev. Lett.\/} \textbf{100}, 066403 (2008).

\bibitem{Woh08} K. Wohlfeld, A.M. Ole\'s, M. Daghofer and P. Horsch,
                   {\em Acta Phys. Polon. A\/} \textbf{115}, this issue (2008).

\bibitem{Dag07} M. Daghofer and A.M. Ole\'s,
                   {\em Acta Phys. Polon. A\/} \textbf{111}, 497 (2007).

\bibitem{Fuj05} J. Fujioka, S. Miyasaka and Y. Tokura,
                   {\em Phys. Rev. B\/} \textbf{72}, 024460 (2005).

\bibitem{Fuj08} J. Fujioka, S. Miyasaka and Y. Tokura,
                   {\em Phys. Rev. B\/} \textbf{77}, 144402 (2008).

\bibitem{Cha06} J. Chakhalian, J.W. Freeland, G. Srajer, J. Strempfer,
                   G. Khaliullin, J.C. Cezar, T. Charlton, R. Dalgliesh,
                   C. Bernhard, G. Cristiani, H.U. Habermeier and B. Keimer,
                   {\em Nat. Phys.\/} \textbf{2}, 244 (2006).

\bibitem{Cha08} J. Chaloupka and G. Khaliullin,
                   {\em Phys. Rev. Lett.\/} \textbf{100}, 016404 (2008).

\bibitem{Mac99} F. Mack and P. Horsch,
                   {\em Phys. Rev. Lett.\/} \textbf{82}, 3160 (1999).



\end{thebibliography}
\end{document}